\documentclass[
 reprint,
 amsmath,amssymb,
 aps,
]{revtex4}
\usepackage[section]{placeins}
\let\Oldsection\section
\renewcommand{\section}{\FloatBarrier\Oldsection}

\let\Oldsubsection\subsection
\renewcommand{\subsection}{\FloatBarrier\Oldsubsection}

\let\Oldsubsubsection\subsubsection
\renewcommand{\subsubsection}{\FloatBarrier\Oldsubsubsection}
\usepackage{graphicx}
\usepackage{dcolumn}
\usepackage{bm}
\makeatletter
\AtBeginDocument{%
  \expandafter\renewcommand\expandafter\subsection\expandafter{%
    \expandafter\@fb@secFB\subsection
  }%
}
\makeatother

\begin{document}

\title{$SU(2)$ Dynamics and Logic machines--part II\footnote{This is part two of the paper\cite{2016arXiv161203276H}}} 

\author{Dawit Hiluf Hailu}
\email{dawit.hailu@mail.huji.ac.il}
\affiliation{Physics Department, Mekelle University, P.O.Box 231, Mekelle, Ethiopia.}
\date{\today}
\begin{abstract}
Solution of quantum dynamics systems can be represented geometrically. Such geometric representations of solutions provides intuitive physical insights. To which end studying dynamics of Quantum systems via $su (2)$ Lie algebra proves to be convenient way of obtaining geometric solution for two level systems. In this paper link is established between two formalisms that made use of Lie algebra to describe equation of motion for quantum system. In both approaches  the Hamiltonian and the density matrix are expressed as a linear combination of the Lie group. To exemplify the approach we consider a very well studied two level system whose coupling laser pulse have area of $\frac{\pi}{2}$. Beyond establishing link between these two formalism we provide analytical solutions using Magnus expansion by exploiting the Sylvester formula where knowledge of eigenvalues suffice to obtain solution. Furthermore we also obtained two constants of motion by assuming time dependent detuning whose time. Consequently we have shown how one can have two disjoint subspaces whose evolution vector is independent of each other. As an application we present two classical binary logic machines, namely CNOT- gate and parity checker, that the two level system's dynamics mimics.  

\end{abstract}

\maketitle



\section{Introduction}
This paper is a continuation of previous work done by the author \cite{2016arXiv161203276H}, for convenience we reproduce most part of the work here. Using the results of the paper we detail some results using two level system in numerical solutions. With which we provide two examples of logic machines.
  
Given any physical system, one can perform certain "operations" or "transformations" with it, example rotations, translations, scale transformations, conformal transformations, Lorentz transformations. Physical transformations form a group from the mathematical viewpoint. In this paper, we make use of the Lie groups $U(n)$ and $SU(n)$, and their respective Lie algebras, generally denoted by $u(n)$ and $su(n)$. The unitary group is defined by: $U(n)\equiv\left\{ U\in M_{n\times n}(\mathbb{C})/UU^+=U^+U=I\right\}$. The special unitary group is defined by: $SU(n)\equiv\left\{ U\in M_{n\times n}(\mathbb{C})/UU^+=U^+U=I,\det (U)=1\right\}$. The group operation is the usual matrix multiplication. We note also that $U(n)$ and $SU(n)$ are compact Lie groups. As a consequence, they have unitary, finite dimensional and irreducible representations. Moreover $U(n)$ and $SU(n)$ are subgroups of $U(m)$ if $m\geq n$. The unitary group has $n^2$ parameters (its "dimension") whereas the special unitary group has $n^2-1$ free parameters (its "dimension"). This means that the unitary group has a Lie algebra generated by the space of $n^2$ dimensional complex matrices while the special unitary group has a Lie algebra generated by the $n^2-1$ dimensional space of hermitian $n\times n$ traceless matrices. 

A very familiar group in quantum mechanics community is the case $n=2$, this is an important group in physics. It appears in many contexts: angular momentum (both classical and quantum), the rotation group, spinors, quantum information theory, spin networks and black holes, the Standard Model, and many other places. The number of parameters of $SU(2)$ is equal to 3. As generators of the Lie algebra associated to this Lie group, called $su(2)$, we can choose for free 3 any independent traceless (trace equal to zero) matrices. The generators of any Lie group satisfy some algebraic and important relations. In the case of dealing with matrix or operator groups, the generators are matrices or operator themselves. These mathematical relations can be written in terms of (ordinary) algebraic commutators.

Feynman, Vernon, and Hellwarth developed geometric analogue of Rabi solutions, for two-level system, which provided additional physical insights. Such approach has been extended into $N$ level system. In this description the dynamics of  $N$ level system is visualised, without introducing new physics, as a vector in an $N^2-1$-dimensional vector space. This requires use of the $SU(N)$ symmetry to embed the atomic variables in the form of a vector with $N^2-1$ components. The exploitation of Lie algebra for study of dynamics has been extended into $N$-level system by Alhassid-Levine (AL)\cite{alhassid1977entropy,AlhassidLevine} and Hioe-Eberly(HE)\cite{Hioe1981,hioe1982nonlinear,hioe1983dynamic}. In Hioe-Eberly approach, for instance, the state of a three-level system is represented by pseudo-spin vector having 8 real components~(9 if we do not impose Normalization). This description provides an elegant geometrical framework.

The Quantum dynamics are calculated by the density matrix formalism \cite{fano1957description}. In this method  a statistical average over the assembly of the expectation values for the individual ions is included. This in turn yields information about observable macroscopic quantities. The formalism is advantageous in that it greatly simplifies the calculations since it allows one to handle a large number of dynamical variables in a systematic way. The physical significance of the individual density matrix elements depends on the representation with which the matrix is calculated. In a representation where $H_0$ is diagonal, the elements of the density matrix are defined as $\rho_{nm}=\langle\psi_n|\hat\rho|\psi_m\rangle$  where $\hat\rho$ is the density operator.  Recall that the dynamical evolution of an $N$-level atomic system is known to be studied in terms of density matrix $\hat\rho$ which obeys the Liouville equation \cite{levine2011quantum}
\begin{equation}
\begin{aligned}
i\hbar\frac{\partial}{\partial t}\hat\rho=&\left[\hat\rho,\hat H\right]
\label{Liouville}
\end{aligned}
\end{equation}
As by definition $\langle\hat O_{mn}\left(t\right)\rangle=Tr\left(\hat O_{mn} \hat\rho\left(t\right)\right)=\rho_{nm}\left(t\right)$ this implies that the dynamics can be equally investigated by studying $\hat O_{mn}$, where $\hat O_{mn}=|m\rangle\langle n|$. 

Although both paths, AL and HE, lead to same equation of motion for the expectation value of the generators the similarity is not straight forward because of the notation used in both formalism. While the AL method expressed closed Lie algebra as a commutation between the Hamiltonian and the generators, and in HE the closure is given by writing the Hamiltonian as a linear combination of the generators.  Moreover the equations of motion, for the expectation values of the generators, are given for row matrices in AL formalism where as HE provided the equation of motion for the columns. On top of this the background and motivation of both approaches also different. We here present the connection between these two derivation. The Alhassid-Levine formalism (AL) is developed in connection to surprisal analysis while the Hioe-Eberly (HE) aimed to extend the optical Bloch equations to three and more level systems. Optical Bloch equations of two level system are derived using both approaches as an example to complete the discussion. The use of expectation values, as opposed to operators, in deriving equation of motion has advantage in that it, i.e the average value, is a number that is measurable and easily solvable numerically and at times analytically. All this seem to make the two routes as if unconnected. This paper not only  aim to bring this two derivations together but also provide appropriate link which serves as a route from one formalism to another.

This work presents the connection between the two formalism, AL and HE.  To describe the equation of motion of the dynamics with the aid of Lie algebra firstly the atomic variables of a quantum  system are  embedded via $SU(N)$  Lie group yielding a new combination of group.  With this Lie group we construct a vector with $N^2-1$ components ($N$ components if normalization is not taken in to account), for $N$-level quantum system, to obtain the equation of motion describing the $SU(N)$ dynamics. Such approach has been shown to reveal hidden constants of motion \cite{hioe1982nonlinear,hioe1983dynamic}.  

The outline of the paper is as follows:  summary of the AL \& HE formalisms will be provided in Sec~(\ref{sec:ALHE}) before pointing out the link between them, AL and HE, in Sec~(\ref{sec:ReALHE}) following by derivation of the equation of motion in Sec~(\ref{sec:EQOM}) for two level system before. Next in Sec~(\ref{sec:solnconstmtn}) solutions, analytical and numerical, are provided along with discussion of constants of motion and super-evolution operator. Then follows discussion of effect of the environment on the two level system Sec~(\ref{sec:disspSysys}). As an application possible implementation of two logic machines are presented  before we provide concluding remarks in Sec~(\ref{sec:conclusion}).
\section{$SU(N)$ Dynamics}
\label{sec:ALHED}
In this section summary of derivations of both approach will be outlined. Both formalisms make use of the Liouville equation of motion to obtain the equation of motion for the expectation value of the generators. However AL use the commutation relation between the generators and the Hamiltonian while HE expanded the Hamiltonian as a linear combination of the generators. It is to be recalled here that every $N\times N$ density matrix is expressible in terms of a diagonal matrix and an element of of the Special unitary group $SU(N)$, generally speaking this provides $N^2$ real parameters if we do not impose normalization. 
\subsection{Summary of Alhassid-Levine (AL) and Hioe-Eberly (HE) formalisms}
\label{sec:ALHE}
For a $N$ level quantum system, where the Hilbert space is spanned by orthonormal states $|m\rangle,|k\rangle\ldots |n\rangle$. One readily sees that the $N^2$ operators $|i\rangle\langle j|, ~ i,j=1,2 \ldots N^2$ form a complete basis for the linear operators of the system. Using which any operator of the system can is expressible as a linear combination of these $N^2$ operators, with which we can form the desired Lie group, by embedding the atomic variables in them. To this end the Lie group used in both formalisms, i.e the generators $\hat G_\alpha$ (formed via linear combination of the $N^2$ operators), are traceless, orthogonal, Hermitian and obey the following properties \cite{Hioe1981,alhassid1977entropy}
\begin{subequations}
\begin{align}
Tr\left(\hat G_\alpha\hat G_\beta\right)=& 2\delta_{\alpha\beta}\label{TrGs}\\
\left[\hat G_{\alpha}, \hat G_{\beta}\right]=&2i f_{\alpha\beta\gamma} \hat G_\gamma\label{Gcommut}
\end{align}
\label{Gprop}
\end{subequations}
where $\delta_{\alpha\beta}$ is the usual Kronecker delta and $f_{\alpha\beta\gamma}$ is the structure constant which are antisymmetric in all indices. With which the density matrix as well as the Hamiltonian are written as a linear combination of the group generators \cite{Hioe1981,alhassid1977entropy}.
\begin{subequations}
\begin{align}
\hat\rho\left(t\right)=&\frac{\hat I}{N}+\frac{1}{2}\sum_{\alpha=1}^{N^2-1} \langle G_\alpha\rangle\left(t\right)\hat G_\alpha\label{expndRho}\\
\hat H\left(t\right)=&\frac{\hbar}{2}\left[\left(\sum_{\beta}^N\omega_\beta\right)\hat I+\sum_{\alpha=1}^{N^2-1} \Gamma_\alpha\left(t\right)\hat G_\alpha\right]\label{expndHaml}
\end{align}
\end{subequations}
where $\hbar\omega_\beta$ is energy of level $\beta$ and $\hat I$ is the identity operator. The coefficients $\langle G_\alpha\rangle\left(t\right)$ and $\Gamma_\alpha\left(t\right)$, expectation value of the generators and torque --respectively, are given by \cite{Hioe1981,alhassid1977entropy}
\begin{subequations}
\begin{align}
\langle G_\alpha\rangle\left(t\right)=&Tr\left(\hat\rho\left(t\right)\hat G_\alpha\right)\label{expectS}\\
\hbar\Gamma_\alpha\left(t\right)=&Tr\left(\hat H\left(t\right)\hat G_\alpha\right)\label{Torque}
\end{align}
\end{subequations}
Making use of the tracelessness of the generators  along with the  the expansion provided in Eq. \eqref{expndRho} it is to be noted that the trace of the density matrix  is $1$, $Tr(\rho)=1$
In their pioneering work on information theory Alhassid-Levine developed equation of motion for the expectation value of generators the summary of the derivation will be outlined. In the aforementioned scheme \cite{alhassid1977entropy} beginning from the expectation value of an operator, i.e $\langle G_{\alpha}\rangle\left(t\right)=Tr\left(\hat\rho\left(t\right)\hat G_\alpha\right)$, and taking derivative of both sides with respect to time, and making use of the Liouville equation of motion for density matrix one arrives at
\begin{equation}
\begin{aligned}
\frac{d}{dt}\langle G_{\alpha}\rangle=& Tr\Big(\frac{1}{i\hbar}[\hat H,\hat\rho]\hat G_{\alpha}\Big)
\label{vecG4AL0}
\end{aligned}
\end{equation}
Using cyclic property of trace and noting that the Hamiltonian is closed under Lie algebra with the generators
\begin{equation}
\begin{aligned}
\left[\hat H,\hat G_\alpha\right]=i \hbar\sum_\beta \hat G_\beta g_{\beta\alpha}
\end{aligned}
\label{HcomGAL}
\end{equation} 
one readily obtains the  equation of motion for the expectation value of generators,  given to be 
\begin{equation}
\begin{aligned}
\frac{d}{dt}\langle G_{\alpha}\rangle=&-\sum_{\beta}\langle G_{\beta}\rangle g_{\beta\alpha}
\label{vecG4AL}
\end{aligned}
\end{equation}
where $\alpha,~\beta=1,2,\dots,N^2-1$, $\langle G_{\beta}\rangle$ is row, and the $g_{\beta\alpha}$ are (possibly complex) numerical coefficients. 

On the other hand, in HE formalism  the Liouville equation of motion for the density matrix is multiplied, on both sides by generators and then trace, $Tr$, is taken to yield \cite{Hioe1981} :
\begin{equation}
\begin{aligned}
i\hbar\frac{d}{dt}\langle G_{\alpha}\rangle=& Tr\Big([\hat H,\hat\rho]\hat G_{\alpha}\Big)\\
=& Tr\Big(\frac{\hbar}{2}\Gamma_{\beta}\left(t\right)[\hat G_{\alpha},\hat G_{\beta}]\hat\rho\Big)
\label{vecG4HE0}
\end{aligned}
\end{equation}
Where the cyclic property of trace and expansion of Hamiltonian as given in Eq.\eqref{expndHaml} has been made use of. Note also that the Hamiltonian commutes with Identity. Next, with the aid of the commutation relation Eq.\eqref{Gcommut} and \eqref{expectS}, i.e the expectation value of the generators, one readily arrives at the equation of motion for the expectation values of the generators 
\begin{equation}
\begin{aligned}
\frac{d}{dt}\langle G_{\alpha}\rangle=&\Gamma_{\beta}\left(t\right)f_{\alpha\beta\gamma}\langle G_{\gamma}\rangle 
\label{vecG4HE}
\end{aligned}
\end{equation}
Therefore as can be seen from,  Eq.\eqref{vecG4AL} and Eq.\eqref{vecG4HE}, of AL and HE respectively, at glance seems unrelated as the notation used are slightly different. It is the intention of the next section to establish the link between these two routes. 
\subsection{Connection between Alhassid-Levine (AL) and Hioe-Eberly (HE) formalisms}                                                                             
\label{sec:ReALHE}
In the AL formalism, Eq. \eqref{vecG4AL} is equation of motion for the generator where $\langle G_{\beta}\rangle$ is row. So to make comparison with HE formalism we need to write them as equation of motions for column $\langle G_{\beta}\rangle^T$, in doing so we should note that due to antisymmetric nature of the numerical coefficients the transpose of the $g$ matrix takes the form $g_{\alpha\beta}=-g_{\beta\alpha}$
\begin{equation}
\begin{aligned}
\frac{d}{dt}\langle G_{\alpha}\rangle=&g_{\alpha\beta} \langle G_{\beta}\rangle 
\label{vecG4ALc}
\end{aligned}
\end{equation}
where indices represent sum over. The comparison of Eq. \eqref{vecG4ALc} with that of HE, Eq. \eqref{vecG4HE}, yields that
\begin{equation}
\begin{aligned}
g_{\beta\alpha}=&\Gamma_{\gamma}\left(t\right) f_{\gamma\alpha\beta}
\label{AL2HE}
\end{aligned}
\end{equation}
Thus the matrix $g$ of AL is the product between the torque vector and antisymmetric structure constant $f_{\alpha\beta\gamma}$ of HE.
\section{Example: Equation of motions for two level system }                                                                             
\label{sec:EQOM}
We now turn to the physical application of the $SU(2)$  groups and $su(2)$ algebras. To see example of the approach we here take a simple  and familiar two level system and describe the dynamics using $SU(2)$ group. The two level system we take is very well studied. We consider a two level system  where  a laser pulse perturbs the systems. For the purpose at hand all we need to know is its Hamiltonian to obtain the equation of motion for the expectation value of the generators. The Hamiltonian we consider is under Rotating Wave Approximation (RWA) and in the interaction picture is known to be \cite{Bergmann2001,shore2008}.
\begin{equation}
\hat H \left(t\right)=\frac{\hbar}{2}\begin{pmatrix}
0 & \Omega  \left(t\right)\\
\Omega \left(t\right) & 2\Delta \\
 \end{pmatrix} 
\end{equation}
where $\frac{\hbar\Omega \left(t\right)}{2}$ is  half-Rabi frequency of the pulse applied and $\hbar\Delta$, i.e detuning, is the difference between the laser frequency and Bohr frequency.  

 For a two-level quantum system, where the Hilbert space is spanned by two orthonormal states $|0\rangle$ and $|1\rangle$. One readily sees that the four operators $|i\rangle\langle j|, ~ i,j=0,1$ form a complete basis for the linear operators of the system in the sense that any linear operator of the system can be written as a linear superposition (with complex coefficients which can be time dependent) of these four operators. It therefore is possible to chose the Pauli matrices, along with the identity operator $\hat I=|0\rangle\langle 0|+|1\rangle\langle 1|$, as our generators, and they are given by \cite{MichaelChuang2010, Eberly1975}
\begin{equation}
\begin{aligned}
 \hat G_1=\begin{pmatrix}
0 & 1\\
1 & 0 \end{pmatrix},       & & \hat G_2=\begin{pmatrix}
0 & -i\\
i & 0 \end{pmatrix} ,      & & \hat G_3=\begin{pmatrix}
1 & 0\\
0 & -1 
 \end{pmatrix}  
 \end{aligned}
 \label{gene}
\end{equation}
The structure constant $f_{\alpha\beta\gamma}$ in this case is the Levi-Civita $\epsilon_{\alpha\beta\gamma}$.  Moreover the non-vanishing components of the fully antisymmetric tensor $\epsilon$ of the constants of $SU(2)$ group given by 
\begin{equation}
\begin{aligned}
\epsilon_{123}=&\epsilon_{231}=\epsilon_{312}=1\\
\epsilon_{132}=&\epsilon_{213}=\epsilon_{321}=-1
\end{aligned}
\label{structSU2}
\end{equation}
To explore the dynamics of the two level using the both scheme, we here express our Hamiltonian, to within an addition of multiple of an identity matrix which commutes with the Hamiltonian, in terms of the generators of the $SU (2)$ group Eq.\eqref{gene} as \cite{Cahn1984} 
\begin{equation}
\begin{aligned}
\hat H\left(t\right)=\frac{\hbar}{2}\Big[\Omega\left(t \right)\hat G_1-\Delta\hat G_3\Big]+const
\end{aligned}
\label{ham2}
\end{equation} 
where $const=\frac{\hbar}{2}\Big(\sum_{\beta=1}^2\omega_\beta\Big)\hat I$, $\hbar\omega_\beta$ is energy of level $\beta$. Note here that, with the exception of a term proportional to $\hat I$, which is constant and thus commutes with any linear operator of the system, the Hamiltonian is a linear combination with real coefficients of the three $su(2)$ generators. This means that the dynamical algebra of the two-level atomic system is $su(2)$. 

Next the equation of motion of a coherence vector whose elements  are the expectation value of the generators as given by Eq.\eqref{expectS} formed to be  $\vec G=\left(\langle G_1\rangle\left(t\right),\langle G_2\rangle\left(t\right),\langle G_3\rangle\left(t\right)\right)^T$ is obtained.

In the AL formalism one is able to find the matrix elements $g_{\beta\alpha}$ using the commutation relation in Eq.\eqref{HcomGAL} and Eq.\eqref{ham2} along with the commutation relation between the generators. For instance we note that the commutation between the Hamiltonian and the generator $\hat G_2$ is 
\begin{equation}
\begin{aligned}
\left[\hat H\left(t\right), \hat G_{2}\right]=& i \hbar \hat G_\beta g_{\beta 2}\\
i \hbar \left( \Omega\left(t\right) \hat G_{3} + \Delta\hat G_{1}\right)=& i \hbar \hat G_\beta g_{\beta 2}\\
\end{aligned}
\label{HcomG2}
\end{equation}
where the indices means sum over. From the last line of Eq.\eqref{HcomG2} it follows that 
\begin{equation}
\begin{aligned}
g_{32}=&\Omega\left(t\right),  && g_{12}=\Delta
\end{aligned}
\label{gx2}
\end{equation}
In like manner one is able to find all the $g_{\beta\alpha}$. Using this knowledge and Eq.\eqref{vecG4AL} the equation of motion, say for $\langle \dot G_2\rangle$, is obtained as
\begin{equation}
\begin{aligned}
\frac{d}{dt}\langle G_2\rangle\left(t\right)=&-\langle G_\beta \rangle g_{\beta2}\\
=&-\Delta\langle G_1 \rangle -\Omega\left(t\right)\langle G_3 \rangle
\end{aligned}
\label{dG2dt}
\end{equation}
To find the equation of motion using HE formalism, we note the torque vector to be $\vec\Gamma\left(t\right)=(\Omega\left(t\right), 0,  -\Delta)^T$, equipped with this and Eq.\eqref{vecG4HE} it is possible to reproduce the previous result as follows
\begin{equation}
\begin{aligned}
\frac{d}{dt}\langle G_2\rangle\left(t\right)=&\Gamma_\beta\left(t\right)f_{2\beta\gamma}\langle G_\gamma \rangle\\
=&-\Delta\langle G_1 \rangle -\Omega\left(t\right)\langle G_3 \rangle
\end{aligned}
\label{dG2dtHE}
\end{equation}

Hence using either of the approach with as given in Eq.\eqref{vecG4AL} and Eq.\eqref{vecG4HE}  the  equation of motion for the coherence vector can readily obtained. 
\begin{equation}
\begin{aligned}
\frac{d}{dt}\vec G=&g\vec G
\label{sdot}
\end{aligned}
\end{equation}
 where $g$ is a $3\times 3$ antisymmetric matrix given by 
\begin{equation}
\begin{aligned}
g=&
 \begin{pmatrix}
  0 & \Delta  & 0 \\
  -\Delta & 0 & -\Omega\left(t\right) \\
  0 & \Omega\left(t\right) & 0
 \end{pmatrix}
\end{aligned}
\label{matrixg}
\end{equation} 
\section{Solution and constants of motion}
\label{sec:solnconstmtn}
One advantage of working in Lie algebra scheme is the ease of obtaining solution for Hamiltonian closed under Lie algebra. In this section we provide the numerical and analytic solution for the two level system introduced in the preceding section. The analytic solution is achieved by use of the Magnus expansion (up to third order) wherein we have employed the knowledge of eigenvalues to obtain the evolution matrix via Sylvester formula.  We  exploit the properties of the $su(2)$ algebra in the formulation of the equations of motion of the dynamical variables of an two-level system, and how one can obtain conservation laws (constants of motion) corresponding to $SU(2)$. 

\subsection{Magnus Approximations}
\label{sec:Magnus}
One way of solving equation Eq.\eqref{sdot} is given by the Magnus Approximation \cite{Magnus1954}.The Magnus expansion, named after Wilhem Magnus, provides an exponential representation of the solution of a first order linear homogenous equation for linear operator. Given an $N\times N$ coefficient matrix $g\left(t\right)$ we wanted to solve the initial value value problem associated with the linear ordinary differential equation, which in our case is the equation of motion for the coherence vectors given by, along with its initial condition:
\begin{equation}
\begin{aligned}
\frac{d \vec{G}\left(t\right)}{dt}=&g\left(t\right)\vec G \left(t\right) &     \vec G \left(0\right) =G_0
\end{aligned}
\end{equation}
Solution of which can be written as 
\begin{equation}
\begin{aligned}
\vec G\left(t\right)=&R\left(t,t_0\right)\vec G\left(t_0\right)
\end{aligned}
\label{su2gsol}
\end{equation}
where $\vec G$ is a 3-dimensional column vector and $R$ is a $3\times3$ super-evolution matrix.
We now make the assumption that the laser is on resonance or close to resonance, meaning $\Delta=0$ or is negligibly small and  consequently the matrix $g$ now commutes with itself at different times, that is $\left[g\left(t_1\right),g\left(t_2\right)\right]=0$. 
The approach proposed by Magnus to solve the matrix initial value problem is to express the solution of the exponential of a certain $N\times N$ function $P\left(t,t_0\right)$
\begin{equation}
\begin{aligned}
\vec{G}\left(t\right)=&e^{P\left(t,0\right) }\vec G \left(0\right)
\end{aligned}
\end{equation}
which is subsequently written as a series expansion
\begin{equation}
\begin{aligned}
P\left(t,0\right)=&\sum_{k=1}^\infty P_k\left(t,0\right)
\end{aligned}
\end{equation}
writing $P\left(t,0\right)=P\left(t\right)$ for simplicity, the first three series reads thus 
 \begin{equation}
\begin{aligned}
P_1\left(t\right)=&\int_{0}^t g\left(t_1\right) dt_1\\
P_2\left(t\right)=&\frac{1}{2}\int_{0}^t dt_1 \int_{0}^{t_1} \left[g\left(t_1\right), g\left(t_2\right)\right] dt_2\\
P_3\left(t\right)=&\frac{1}{6}\int_{0}^t dt_1 \int_{0}^{t_1} dt_2 \int_{0}^{t_2} \left[g\left(t_1\right),\left[g\left(t_2\right), g\left(t_3\right)\right]\right]dt_3\\
+&\frac{1}{6}\int_{0}^t dt_1 \int_{0}^{t_1} dt_2 \int_{0}^{t_2}\left [g\left(t_3\right),\left[g\left(t_2\right), g\left(t_1\right)\right]\right]  dt_3
\end{aligned}
\label{Gseries}
\end{equation}
where $\left[g_1,g_2\right]=g_1g_2-g_2g_1$.  In search of better approximation one needs to include more terms in the Magnus expansion. In this paper we will be considering only  the first three series $P_k\left(t\right), k=1, 2, 3 $, therefore the solution for Eq.\eqref{sdot} now becomes
\begin{equation}
\begin{aligned}
\vec G\left(t\right)=&R\left(t,t_0\right)\vec G\left(t_0\right)\\
\vec{G}\left(t\right)=&e^{P^{(3)}\left(t\right)}\vec G \left(0\right)
\end{aligned}
\label{Soft}
\end{equation}
where $P^{(3)}\left(t\right)$ is sum of the three series given in Eq.\eqref{Gseries} and we set initial time to be at zero $t_0=0$. Thus the third order Magnus expansion yields the desired superevolution matrix $R\left(t,t_0\right)$. One way of determining $e^{P^{(3)}}$ is using the Sylvester formula. The Sylvester formula is a way of solving any exponential function by making use of eigenvalues \cite{19dubiousways,Tarantola}. To this end let $\gamma_j$  be an eigenvalue of $P^{(3)}\left(t\right)$, we thus can write the exponent using Sylvester formula (iff we have distinct eigenvalues) as  
\begin{equation}
\begin{aligned}
e^{P^{(3)}\left(t\right)}=&\sum_{j=1}^{3}e^{\gamma_j}\prod_{j\neq k=1}^{3}\frac{P^{(3)}\left(t\right)-\gamma_j  I}{\gamma_k-\gamma_j}
\end{aligned}
\label{sylvstr}
\end{equation}
The third order Magnus is  obtained to be of the form  
\begin{equation}
\begin{aligned}
P^{(3)}\left(t\right)=&\begin{pmatrix}
 0 &  \eta\left(t\right)  & \lambda_0\left(t\right) \\
  -\eta\left(t\right) & 0 & -\zeta\left(t\right) \\
  -\lambda_0\left(t\right) &  \zeta\left(t\right) & 0
 \end{pmatrix}
\end{aligned}
\label{Hintg}
\end{equation}
where $\eta\left(t\right)=\Delta'\left(t\right)+\lambda_1\left(t\right)$ and $\zeta\left(t\right)=\Omega'\left(t\right)+\lambda_2\left(t\right)$ with $\Omega' \left(t\right)=\int_{0}^t \Omega\left(t_1\right) dt_1$, $\Delta'\left(t\right)=\int_{0}^t \Delta  dt_1$ and
\begin{equation}
\begin{aligned}
\lambda_0\left(t\right)=&\frac{\Delta}{2}\int_{0}^t dt_1\int_{0}^{t_1} dt_2\left(\Omega_1-\Omega_2\right)\\
\lambda_1\left(t\right)=&-\frac{\Delta}{6}\int_{0}^t dt_1 \int_{0}^{t_1} dt_2 \int_{0}^{t_2} dt_3\\ &\times\left(\Omega_1\left(\Omega_2-2\Omega_3\right)+\Omega_2\Omega_3\right)\\
\lambda_2\left(t\right)=&-\frac{\Delta^2}{6}\int_{0}^t dt_1 \int_{0}^{t_1} dt_2 \int_{0}^{t_2} dt_3\left(\Omega_1-2\Omega_2+\Omega_3\right)
\end{aligned}
\label{integrands}
\end{equation}
where we used the notation $\Omega_j=\Omega\left(t_j\right)$ for simplicity.
In what follows  we omit the time argument $\left(t\right)$ unless it is needed for clarity. The eigenvalues of $P^{(3)}$, are readily obtained to be $\{0, -i\xi, i\xi\}$, where $\xi=\sqrt{\lambda^2+\zeta^2+\eta^2}$. It is worth pointing out here that we have distinct eigenvalues. 
Therefore we can now express our solution for the coherence vector in terms of the Sylvester formula. To this end, making use of the eigenvalues and  Eq.\eqref{Soft} along with Eq.\eqref{sylvstr}  and noting that the system is initially prepared to be on the ground state,  we find the following solution 
\begin{equation}
\begin{aligned}
\vec{G}\left(t\right)=&
 \begin{pmatrix}
 -\frac{\zeta\eta-\zeta\eta\cos{\xi}-\lambda\xi\sin{\xi}}{\xi^2}\\
  \frac{-\lambda\eta+\lambda\eta\cos{\xi}-\zeta\xi\sin{\xi}}{\xi^2}\\
 \frac{\eta^2+(\lambda^2+\xi^2)\cos{\xi}}{\xi^2}
 \end{pmatrix}
\end{aligned}
\label{st1}
\end{equation}
If however we consider only the first term in the Magnus expansion, it follows that $\lambda=\lambda_1=\lambda_2=0$, and  accordingly we have to modify our solution-- Eq.\eqref{st1} to be
\begin{equation}
\begin{aligned}
\vec{G}\left(t\right)=&
 \begin{pmatrix}
 \frac{\Delta'\Omega'}{\xi^2}\big(-1+\cos\xi\big)\\
 - \frac{\Omega'}{\xi}\sin\xi\\
 \frac{\Delta'^2}{\xi^2}+\frac{\Omega'^2}{\xi^2}\cos\xi
 \end{pmatrix}
\end{aligned}
\label{st2}
\end{equation}
where now $\xi=\sqrt{\Delta'^2+\Omega'^2}$
\subsection{Results and discussion}
\label{sec:results}
Now that the relevant equations necessary to proceed are complete, we here make  comparison of the analytical and numerical solutions.
It is worth pointing out here that because of the pulse area we used, i.e. $\frac{\pi}{2}$ we do not see the Rabi oscillations, we instead see that, at the end of the pulse interaction, we are able to create  superpositions between states $|0\rangle$ and $|1\rangle$. To see the Rabi oscillation, where in the population fluctuates between states $|0\rangle$ and $|1\rangle$, one needs to use a laser whose pulse area is an integral multiple of $\pi$.  

To begin with the case of exact resonance where $\Delta=0$ is solved first, in this case as can be seen from Eq.\eqref{integrands} that all the $\lambda_k$ terms vanish,  because all the  terms are dependent on $\Delta$, and consequently the Hamiltonian commutes with itself. Therefore the solutions obtained using Magnus expansion and numerical solution are identical as shown in Fig.(\ref{Soln123ndOMagnusD0}i). But if we have a non-zero detuning but small value the results obtained numerically and that using Magnus expansion may differ. We see in Fig.(\ref{Soln123ndOMagnusD0}ii-a) comparison of solutions obtained via numerical and first order Magnus. As we  took a very  small value of detuning  we have a very small value resulted from the commutation of the Hamiltonian with itself at different times. Consequently the solution we obtained using the first order Magnus approximation is close to the numerical but does not fully agree with the numerical solution. 
\graphicspath{{Figures//}}
\begin{figure*}[htp]
\centering
(i)\includegraphics[width=2.3 in]{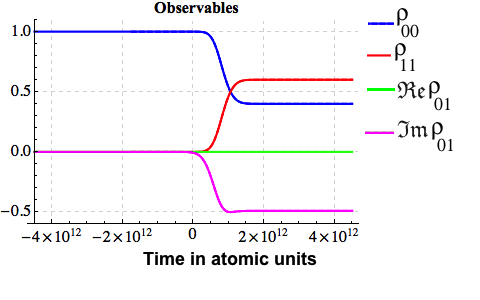}
(ii)\includegraphics[width=4.3 in]{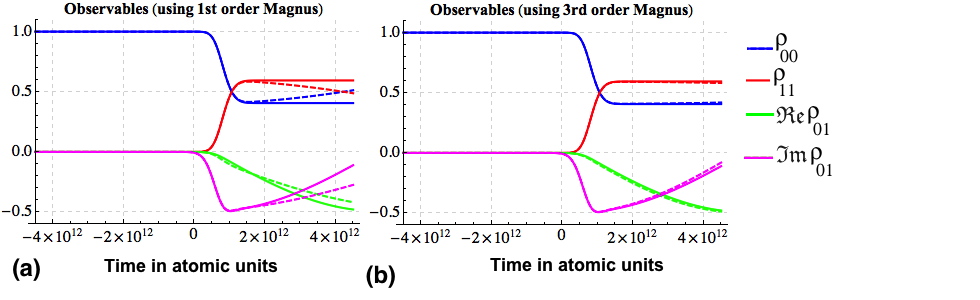}
\label{Soln123ndOMagnusD0}
\caption{(Color online) (i) Solution using first order Magnus, i.e. when $\lambda=\lambda_1=\lambda_2=0$ and $\Delta=0$.  (ii-a) Solution using first order Magnus, i.e. when $\lambda=\lambda_1=\lambda_2=0$ and $\Delta\neq0$.~(ii-b) third order Magnus, i.e. when $\lambda=\lambda_1=\lambda_2\neq0$. The solid lines are numerical solution whereas the dashed lines are analytical solution. Key: blue=$\rho_{00}$, red=$\rho_{11}$, green=Real part of $\rho_{01}$, and magenta=Imaginary part of $\rho_{01}$ }
\label{Soln123ndOMagnusD0}
\end{figure*}
Whereas if we use the third order Magnus expansion as it means we include more correction terms the  the solution is seen to be improved Fig.(\ref{Soln123ndOMagnusD0}ii-b)
\subsection{Constants of Motion}
\label{sec:constmtn}
States are labeled by specific values of their properties, which do not change with time - these properties are called constants of motion. A constant of motion in quantum mechanics is defined as an observable-operator that commutes with the Hamiltonian. In two-level atoms (and also in N-level atoms), constants of motion are conveniently discussed in terms of the Bloch sphere and restrictions for the Bloch vector moving on the Bloch sphere. In this case, the underlying symmetry group that can explain the constants of motion is $SU(N)$. The Hamiltonian of the N-level atom can be expressed in terms of the group generators of $SU(N)$, and the dynamical variables of the system can be associated with these generators. Hioe and Eberly have shown how to exploit the properties of the $su(N)$ algebra in the formulation of the equations of motion of the dynamical variables of an N-level system, and how one can obtain conservation laws (constants of motion) corresponding to $SU(N)$. 

In the two-level system the generator $\hat I$ commutes with the rest of the generators, i.e $[\hat G_\alpha,\hat I]=0$ -- where $\alpha=1,2,3$. This in turn implies that the generator $\hat I$  does not contribute to the dynamical equations of motion, as it commutes with any linear operator of the system, this means that the generator $\hat I$  is a constant of motion (i.e. $\frac{d\hat I}{dt}=0$). Moreover in search of constants of motion beyond the identity operator let us now assume that the detuning to be time dependent. Furthermore consider the case where both the detuning and the laser pulse to have same time dependence but  different amplitude.
\begin{equation}
\begin{aligned}
\Omega\left(t\right)=&\Omega_0 q\left(t\right)\\
\Delta\left(t\right)=&\Delta_0 q\left(t\right)
\end{aligned}
\label{DeltaOmega}
\end{equation}
with $q\left(t\right), \Omega_0, \Delta_0$ the time dependence function,  peak values of the laser pulse, and peak value of the Detuning respectively. Introducing $\epsilon\left(t \right)=\sqrt{\Omega^2+\Delta^2}$ and thus following footsteps of \cite{PhysRevA.11.1641} this would equip us with dimensionless parameters $\frac{\Omega}{\epsilon},\frac{\Delta}{\epsilon}$. Armed with this new information let us now make transformation of the generators $\hat G_\alpha$, of Eq.\eqref{gene}, into new generators labelled as $\hat F_\alpha$
\begin{equation}
\begin{aligned}
\hat F_1=&\frac{\Omega}{\epsilon} \hat G_1-\frac{\Delta}{\epsilon} \hat G_3\\
\hat F_2=&\hat G_2\\
\hat F_3=&\frac{\Delta}{\epsilon} \hat G_1-\frac{\Omega}{\epsilon} \hat G_3
\end{aligned}
\label{Fgene}
\end{equation}
Note here that this new generators have same commutation relation as the generators $\hat G_\alpha$. Also notably, using Eqs. \eqref{expndRho} and \eqref{expndHaml}, we can express the density matrix and the Hamiltonian in terms of the new generators, and consequently the Hamiltonian takes the form
\begin{equation}
\begin{aligned}
\hat H\left(t\right)=\hbar\epsilon\left(t \right)\hat F_1+const
\end{aligned}
\label{hamF2}
\end{equation} 
Equation \eqref{hamF2} reveals two interesting and points, first notice the Hamiltonian is no more dependent on the generators $\hat F_3$, compare with Eq.\eqref{ham2}, because the Hamiltonian now commutes with the generator $\hat F_1$, i.e. $[\hat H\left(t\right),\hat F_1]=0$, we conclude $\hat F_1$ is a constant of motion and it thus follows that
 \begin{equation}
\begin{aligned}
\frac{d}{dt}\hat F_1=0
\end{aligned}
\label{dF1dt}
\end{equation} 
Following the steps outline in section \ref{sec:EQOM}, and using the notation $\vec F=(F_1,F_2,F_3)^T$, where $F_\alpha$ is the expectation value of the generators, one readily obtains the dynamical equations of motion for the new generators to be
\begin{equation}
\begin{aligned}
\frac{d}{dt}\vec F=&f\vec F
\end{aligned}
\label{dFdt1}
\end{equation}  
where 
\begin{equation}
\begin{aligned}
f=&\begin{pmatrix}
f_{1\times 1} & 0\\
0  & f_{2\times2}\end{pmatrix}, \\
f_{1\times 1}=&\begin{pmatrix} 0 \end{pmatrix}, &
f_{2\times2}=&\begin{pmatrix}
0 & -\epsilon\\
\epsilon  & 0\end{pmatrix}
\end{aligned}
\label{dFdt2}
\end{equation} 
This informs us that the newly transformed vector $\vec F$ is decomposed into two vectors that live in separate subspaces  and consequentially yields two more constants of motion
\begin{equation}
\begin{aligned}
F_1\left(t\right)^2=&const,\\
F_2\left(t\right)^2+F_3\left(t\right)^2=& const
\end{aligned}
\label{dFdtconst}
\end{equation} 
Physically the last equation means that the length of $\vec F_1$ and $\vec F_{23}=(F_2,  F_3)^T$ is separately conserved, in addition to this, as usual, the total length of the vector $\vec F$ is also conserved~\cite{Eberly1975,hioe1982nonlinear,hioe1983dynamic},
\begin{equation}
\begin{aligned}
F_1\left(t\right)^2+F_2\left(t\right)^2+F_3\left(t\right)^2=& const
\end{aligned}
\label{Fdtconst}
\end{equation} 
It is worth pointing out that for exact resonance, i.e. $\Delta=0$, one readily notes that $\hat F_\alpha=\hat G_\alpha$ and because now $[H, G_1]=0$ at exact resonance $G_1$ is new constant of motion. Setting $\Delta=0$ shows that now the dynamical equation of motion can be decomposed into  two subspaces each of which yield constants of motion given to be
\begin{equation}
\begin{aligned}
G_1\left(t\right)^2=&const,\\
G_2\left(t\right)^2+G_3\left(t\right)^2=& const
\end{aligned}
\label{dgdtconst}
\end{equation} 
From which follows that the length of $\vec G_1$ and $\vec G_{23}=( G_2, G_3)^T$ is separately conserved, in addition to this, as usual, the total length of the vector $\vec G$ is also conserved~\cite{Eberly1975}. 

Another important feature to notice from Eq.\eqref{dFdt1} is that, not only the block diagonal matrix $f\left(t\right)$ commutes with itself at different times, i.e $[f\left(t_1\right),f\left(t_2\right)]=0$, but also the disjoint matrices $f_k\left(t\right), k=1,2$ commutes with thierself at different times, i.e $[f_k\left(t_1\right),f_k\left(t_2\right)]=0$. This means that we can employe the Magnus solution, outlined in section~(\ref{sec:Magnus}), and  obtain the solution for the new vector to be $\vec{F}\left(t\right)=M\left(t,0\right)\vec{F}\left(0\right)$ where $M\left(t,0\right)$ is given to be
\begin{equation}
\begin{aligned}
M\left(t,0\right)=&\begin{pmatrix}
1 & 0 &0\\
0 & \cos\epsilon' &-\sin\epsilon'\\
0 & \sin\epsilon' & \cos\epsilon'
\end{pmatrix}
\end{aligned}
\label{MstF1}
\end{equation}
 where $\epsilon'=\int_{0}^t\epsilon\left(t'\right)dt'$. To this end iff we again assume the system is initially prepared to be on the ground state, this entails to the initial vector being $\vec F\left(0\right)=(-\frac{\Delta_0}{\epsilon_0},0,\frac{\Omega_0}{\epsilon_0})^T$ with $\epsilon_0=\sqrt{\Omega_0^2+\Delta_0^2}$ hence the solution takes the form
\begin{equation}
\begin{aligned}
\vec{F}\left(t\right)=&
 \left(
 -\frac{\Delta_0}{\epsilon_0},
  -\frac{\Omega_0\sin{\epsilon'}}{\epsilon_0},
 \frac{\Omega_0\cos{\epsilon'}}{\epsilon_0}
 \right)^T
\end{aligned}
\label{stF1}
\end{equation}
We can see from the solution Eq.\eqref{stF1} that $F_1\left(t\right)=F_1\left(0\right)$ does not change with time and 
\begin{equation}
\begin{aligned}
|F_1\left(t\right)|^2=&|F_1\left(0\right)|^2\\
|F_2\left(t\right)|^2+|F_3\left(t\right)|^2=&|F_2\left(0\right)|^2+|F_3\left(0\right)|^2
\end{aligned}
\label{FkConserve}
\end{equation} 
To conclude we showed that the time evolution of the pseudospin vector $\vec G$ can be studied in terms of the time evolution of two independent vectors of dimensions one and two, evolving in two separate subspaces of their respective dimensions provided the conditions of Eq.\eqref{DeltaOmega} are met; and/or if system under consideration is on exact resonance (i.e. $\Delta=0$). Following this fact we then provided two constants of motion representing the square of the length of each vector. Moreover the sum of these two vectors is the square of the length of the vector which is commonly known as the constant of motion $\vec G$.
\subsection{Super-Evolution Matrix }
\label{sec:supeEvolWN}
In this section we will explore the super-evolution, where super stands for an operator acting on another operator, matrix that will be obtained via Wei-Norman approach, WN \cite{wei1963lie,wilcox1967exponential}. WN technique tells us that the evolution operator, of a Hamiltonian that is a linear combination of Lie group generators, can be expressed as an ordered product of exponentials, whose arguments are the product of a time-dependent
function and generator of the group. 
Solution of Eq.\eqref{dFdt1} can be formally written as 
\begin{equation}
\begin{aligned}
\vec F\left(t\right)=&M\big(t,0\big)\vec F\left(0\right)
\end{aligned}
\label{transform}
\end{equation}
where $\vec F$ being three dimensional column vector  whose components are the average values of the observables, and the super-evolution matrix $M$ an $3\times 3$ matrix. 
At this point it is helpful relating  the super-evolution matrix $M$ to the evolution operator $\hat U$. To this aim, let us notice that, according to the $3\times 3$ representation of the generators $\hat F_{\alpha}$,  \cite{dattoli1991matrix}
\begin{equation}
\begin{aligned}
\big(\digamma_{\alpha}\big)_{\beta\gamma}=&-\imath f_{\alpha,\beta,\gamma}
\end{aligned}
\label{adjF}
\end{equation}
This representation is called adjoint representation. The explicit matrix elements of the adjoint representation are given by the structure constants of the algebra. One can rewrite the Liouville equation in the following form (for $\hbar=1$)
\begin{equation}
\begin{aligned}
\imath\frac{d}{dt}\rho\left (t\right)=&[H,\rho]\\
\imath\frac{d}{dt}\rho\left (t\right)=&ad_H\left(\rho\right)
\end{aligned}
\label{adjL}
\end{equation}
where $ad_H$ is so-called commutator superoperator, because it is an operator acting on another operator, and is defined to be
\begin{equation}
\begin{aligned}
ad_H=\frac{\hbar}{2}\sum_{\alpha}\Gamma_{\alpha}\left(t\right)ad_{F_{\alpha}}
\end{aligned}
\label{adjH}
\end{equation}
where $ad_{F_{\alpha}}$ is defined, following Eq.\eqref{adjF} to be
\begin{equation}
\begin{aligned}
\big(ad_{F_{\alpha}}\big)_{\beta\gamma}=&-\imath f_{\alpha,\beta,\gamma}
\end{aligned}
\label{adjFalpha}
\end{equation}
On making use of the adjoint representation and the fact that $\langle \mathcal F\left(t\right)\rangle =Tr\big(\hat\rho\left(t\right)\hat  {\mathcal F}\big)$ we now can rewrite the equation of motion for the coherence vector in the following form 
\begin{equation}
\begin{aligned}
\imath\frac{d}{dt}\vec {\mathcal F}=&ad_H\vec  {\mathcal F}
\end{aligned}
\label{adjFch}
\end{equation}
Notice that the passing from $\rho$ in Eq.\eqref{adjL} to $\mathcal F$ in Eq.\eqref{adjFch} is equivalent to passing to the adjoint representation of the Lie algebra $su(N)$. If we now use the notation $ad_H=\mathcal H$ and $ad_{F_{\alpha}}=\mathcal F$ we can rewrite Eq.\eqref{dsdtgs}  in the matrix form as
\begin{equation}
\begin{aligned}
\imath \frac{d}{dt} \mathcal F=&\mathcal H \mathcal F
\label{adjdiF}
\end{aligned}
\end{equation}
Using Eq.\eqref{transform} it turns into the Schr\"{o}dinger equation for the matrix $M\left(t,0\right)$
\begin{equation}
\begin{aligned}
i\frac {d }{dt}M\left(t,0\right)=&\mathcal H M\left(t,0\right), &&M\left(t,0\right)=I\\
\end{aligned}
\end{equation}
Dropping the time argument $M\left(t,0\right)=M$ for simplicity. 
\begin{equation}
\begin{aligned}
i\frac {d M}{dt}M^{-1}=&\mathcal H ,
\end{aligned}
\label{eqMtnM}
\end{equation}
The interpretation of $M$ as the geometrical counterpart of $\hat U$ immediately follows. In the WN approach the evolution matrix is expressed in terms of closed lie algebra; for generators $\hat  {\mathcal F}$ with corresponding time dependent coefficients $\Upsilon\left(t\right)$, the unitary super-evolution is expressed as 
\begin{equation}
\begin{aligned}
\hat M \left(t\right)=&\prod_{\alpha}e^{\Upsilon_{\alpha}\left(t\right)\hat  {\mathcal F}_{\alpha}}\\
\end{aligned}
\label{weiNrmnM}
\end{equation}
where the adjoint representation generators $ {\mathcal F_\alpha}$
\begin{equation}
\begin{aligned}
\mathcal F_1=&\begin{pmatrix}
0 & 0 & 0\\
0 & 0 & -1\\
 0 & 1 & 0 \end{pmatrix}, &&
 \mathcal F_2=&\begin{pmatrix}
0 & 0 & 1\\
0 & 0 & 0\\
 -1 & 0 & 0 \end{pmatrix}, &&
 \mathcal F_3=&\begin{pmatrix}
0 & -1 & 0\\
1 & 0 & 0\\
 0 & 0 & 0 \end{pmatrix} 
\end{aligned}
\label{mathCalF}
\end{equation}
From Eqs.\eqref{eqMtnM}, \eqref{weiNrmnM} and the adjoint generators \eqref{mathCalF} we 
\begin{equation}
\begin{aligned}
\frac {d M}{dt}M^{-1}=&\dot\Upsilon_1\mathcal F_1+\dot\Upsilon_2~e^{\Upsilon_1\mathcal F_1}\mathcal F_2~e^{-\Upsilon_1\mathcal F_1}+\dot\Upsilon_3~e^{\Upsilon_1\mathcal F_1}~e^{\Upsilon_2\mathcal F_2}\mathcal F_3~e^{-\Upsilon_2\mathcal F_2}~e^{-\Upsilon_1\mathcal F_1}
\end{aligned}
\label{eqMtnMc}
\end{equation}
Using the Baker-Campell-Hausrdoff relation 
\begin{equation}
\begin{aligned}
e^A~B~e^{-A}=& B + [A,B]+\frac{1}{2!}[A,[A,B]]+\frac{1}{3!}[A,[A,[A,B]]]+\ldots
\end{aligned}
\label{BCH}
\end{equation}
we readily see that
\begin{equation}
\begin{aligned}
e^{\Upsilon_1\mathcal F_1}\mathcal F_2~e^{-\Upsilon_1\mathcal F_1}=& \mathcal F_2\cos\Upsilon_1+\mathcal F_3\sin\Upsilon_1\\
e^{\Upsilon_1\mathcal F_1}e^{\Upsilon_2\mathcal F_2}\mathcal F_3~e^{-\Upsilon_2\mathcal F_2}~e^{-\Upsilon_1\mathcal F_1}=&\mathcal F_1\sin\Upsilon_2-\mathcal F_2\cos\Upsilon_2\sin\Upsilon_1+\mathcal F_3\cos\Upsilon_2\cos\Upsilon_1
\end{aligned}
\label{BCHre}
\end{equation}
Equating with the Hamiltonian, for the Hamiltonian given in Eq.\eqref{matrixg}, and putting it in matrix form we get (note the Hamiltonian becomes $\mathcal H=\epsilon\mathcal F$ and the RHS will be $(\epsilon,0,0)^T$ for the case given by Eq.\eqref{dFdt1})
\begin{equation}
\begin{aligned}
\begin{pmatrix}
1 & 0 & \sin\Upsilon_2\\
0 & \cos\Upsilon_1 & -\cos\Upsilon_2\sin\Upsilon_1\\
 0 & \sin\Upsilon_1 & \cos\Upsilon_1\cos\Upsilon_2 \end{pmatrix} \begin{pmatrix}
\dot\Upsilon_1\\
\dot\Upsilon_2\\
 \dot\Upsilon_3 \end{pmatrix}=&\begin{pmatrix}
\Omega\\
 0\\
 -\Delta \end{pmatrix} 
\end{aligned}
\label{wenMat}
\end{equation}
In short hand notation we cab write Eq.\eqref{wenMat} as
\begin{equation}
\begin{aligned}
W\dot\Upsilon=&I
\end{aligned}
\label{wenMatshrt}
\end{equation}
So as to obtain solution we demand that $Det |W|=\cos\Upsilon_2\neq0$, and the inverse of $W$ takes the form
\begin{equation}
\begin{aligned}
W^{-1}=&
\begin{pmatrix}
1 & \sin\Upsilon_1\tan\Upsilon_2 & -\cos\Upsilon_1\tan\Upsilon_2\\
0 & \cos\Upsilon_1 & \sin\Upsilon_1\\
 0 & \sec\Upsilon_1\sin\Upsilon_2 & \cos\Upsilon_1\sec\Upsilon_2 \end{pmatrix} 
 \end{aligned}
\label{inverseW}
\end{equation}
From Eq.\eqref{wenMat} we get that
\begin{equation}
\begin{aligned}
\dot\Upsilon_1+\dot\Upsilon_3\sin\Upsilon_2=\Omega\\
\dot\Upsilon_2\cos\Upsilon_1-\dot\Upsilon_3\cos\Upsilon_2\sin\Upsilon_1=0\\
\dot\Upsilon_2\sin\Upsilon_1+\dot\Upsilon_3\cos\Upsilon_2\cos\Upsilon_1=-\Delta\\
\end{aligned}
\label{param}
\end{equation}
which yields
\begin{equation}
\begin{aligned}
\Upsilon_1=&\cot^{-1}\left(\cos\Upsilon_2\frac{\dot\Upsilon_3}{\dot\Upsilon_2}\right)\\
\Delta=&\sqrt{\dot\Upsilon_2^2+\dot\Upsilon_3^2\cos^2\Upsilon_2}\\
\Omega=&\dot\Upsilon_1+\dot\Upsilon_3\sin\Upsilon_2\\
\end{aligned}
\label{paramsol}
\end{equation}
For the case with the block diagonal Hamiltonian $f$ with $\mathcal H=\epsilon\mathcal F$ where we have $(\epsilon,0,0)^T$ in the RHS of Eq.\eqref{wenMat}, the solution yields
\begin{equation}
\begin{aligned}
\dot\Upsilon_1=&\epsilon,  & & \dot\Upsilon_2=\dot\Upsilon_3=0
\end{aligned}
\label{paramfsol}
\end{equation}
The case $\Upsilon_2=0$ reproduces the evolution matrix Eq.\eqref{MstF1}, i.e $W=M$,  with $\Upsilon_1=\epsilon'=\int_{0}^t\epsilon\left(t'\right)dt'$
\begin{equation}
\begin{aligned}
W=&
\begin{pmatrix}
1 & 0 & 0\\
0 & \cos\Upsilon_1 & -\sin\Upsilon_1\\
 0 & \sin\Upsilon_1 & \cos\Upsilon_1 \end{pmatrix}
 \end{aligned}
\label{WeqlsM}
\end{equation}
 \section{Dissipative System }                                                                            
 \label{sec:disspSysys}
So far we have focused our attention on a closed system, which is an isolated from the environment, and consequently does not interact with the outside environment ; and hence it is not affected by the environment. But such assumptions are unrealistic, as it is extremely difficult to completely isolate a system.  All measurements are subject to fluctuations. In this section we will include the effect of the noise, and provide the general form of the dissipation superoperator. The interaction with the environment often causes the system to relax to an equilibrium state that is statistically mixture of its energy eigenstates. Population relaxation happens when the population of the energy eigenstates change, which usually happens due to spontaneous emission or absorption of quanta of energy at random times. 

The state of an $N$ level quantum system is usually represented by a density operator $\rho$ acting on Hilbert space $\mathcal H$. When the system is isolated from its environment then its evolution is given by the quantum Liouville equation Eq.\eqref{Liouville} \cite{levine2011quantum}. However, the dynamics of an open system that is affected by the environment has to account the relaxation terms. The effect of the environment leads to population and phase relaxation (decay and decoherence, respectively), which ultimately causes it to relax to equilibrium state.  Usually we add the effect of the environment phenomenologically. We therefore restrict our attention to the common case where the effect of the environment  leads to population and phase relaxation (decay and decoherence respectively) of the system. 

To account population relaxation as a result of the interaction with the environment we must modify the system's quantum Liouville equation Eq.\eqref{Liouville} to include the effect. Relaxation superoperator \cite{schirmer2004constraints} can always be decomposed into two parts, one accounting for population relaxation and the other for pure phase relaxation (Dephasing). This decomposition yields a general formula for decoherence rates induced by population decay, which is consistence with the physical expectations. Dephasing can be further decomposed into two components, one that originates from the population relaxation, and another that does not. Atomic coherences are subject to dephasing and decoherences\cite{mukamel1999principles}. We use the terms "dephasing" and "decoherence" to distinguish between two different effects of phase changes in atomic coherences. The former is related to inhomogeneous broadening while the latter is related to homogenous broadening. Both dephasing and decoherence destroy the encoded information. 

Population relaxation necessarily induces phase relaxation, \cite{schirmer2004constraints} and we will later derive the explicit expressions for the contribution of population relaxation for the phase relaxation rates. When the interaction of the system with the environment destroys the phase correlation between quantum states, and consequently converts coherent superposition states into incoherent mixed states, we say Phase relaxation occurred. Since the decoherence determined by the off-diagonal elements in our expansion of the density operator this effect can be modelled as the decay of the off-diagonal elements of $\rho$ 
We now convert the above dissipation matrix into a dissipation matrix for the coherence vector $\vec G=(G_1,G_2,G_3)^T$. Collecting the relaxation terms of the coherences $G_1,G_2$, and including the dynamics for the isolated system,i.e non-relaxation terms, we find the following equation of motion
\begin{equation}
\begin{aligned}
\frac{d}{dt}\begin{pmatrix}
G_{1}\\
G_{2}\\
G_{3} \end{pmatrix}=\begin{pmatrix}
-\Gamma_{01} & \Delta-\Gamma'_{01} & 0 \\
-\Delta+\Gamma'_{01} & -\Gamma_{10} & -\Omega \\
 0 & \Omega & -2\gamma_{01}
\end{pmatrix}
\begin{pmatrix}
G_{1}\\
G_{2}\\
G_{3}
\end{pmatrix}
\end{aligned}
\label{reLaxs}
\end{equation}
note that $\gamma_{01}=\gamma_{10}$

\section{Proposed Logic Machines}
\label{sec:LogMachines}
To begin with it is worth stating that the logic machines discussed in this manuscript are classical binary logic machines, and the physical system considered is quantum system. In this line of thinking \cite{remacle2008inter} a physical quantum system is shown to act as either an ON/OFF switch \cite{Kompa:2001aa} and/or was proved to perform complicated calculations like that of full addition and subtraction\cite{remacle2006all}. 
Because we use the same logic assignment for both logic machines we propose, we here outline the logic assignments along with the threshold values. The logic assignment for pulse is \textbf{0} if the pulse is \textbf{OFF} and logic value \textbf{1} if the pulse is \textbf{ON}. For the initial state the following logic assignment are made: if all the populations are in state $|0\rangle$ the corresponding logic value is \textbf{0}, and if it is in state $|1\rangle$ the corresponding logic assignment is \textbf{1}. Because we have two inputs (i.e, the pulse applied and initial state), and because each can be either \textbf{0} or \textbf{1}, we have $2^2=4$ possible combination of inputs as seen in the truth table of Table~(\ref{table:CNOT}). The logic assignment for the final populations is to be logic \textbf{1} if we have reading of populations $\geq 0.6$ else it is \textbf{0}. That means if we have a population of $\geq 0.6$ at state $|m\rangle$, after the system interacts with the laser pulse, then the corresponding logic is assigned to be $m$ where $m=0,1$. In like manner for the coherences it is logic \textbf{1} if the absolute value of the coherences is $ \geq 0.5$, else \textbf{0}.
\subsection{Controlled NOT gate (CNOT)}
\label{sec:CNOT}
So far the dynamics of the two level system at hand has been studied on top of developing analytical solutions which we made comparison with the numerical solutions and found them to fit in using $3^{rd}$ order expansion. In this section attempt is made to correlate the dynamics and CNOT gate operation. The superevolution operator that is obtained using Sylvester formula is propagating the system from initial time $t_0$ to some final time $t$, this coupled with the profile of the laser pulse used the dynamics is seen to flip one state to another similar to that of CNOT gate operation.  
The Controlled-NOT gate, CNOT, takes as an input two bits (a control bit and a target bit) and performs the following operation: If the control bit is set to zero it does nothing, if it is set to one, the target bit is flipped.  The CNOT gate is that is an basic element in the building quantum computer. 

We here discuss the mimicry of the dynamics of the two level system to the behavior of CNOT gate. To this end recall that the system is initially prepared to be on the ground state $|0\rangle$. From which we distribute the population between the ground and excited state by applying a laser pulse. In doing so and with appropriate profile of the pulse we are able to create coherences. To make use of the dynamics for logic we have to first assign our input/output variables. The input variable we consider are the pulse applied and initial state of the system; and as our out put we read coherences between the two states, $|0\rangle\leftrightarrow|1\rangle$ and the final reading of the population at each state. Next we have to assign the corresponding logic value, is it  \textbf{0} or  \textbf{1}, as discussed below.

\begin{table}[!h]
\centering
\begin{tabular}{ | l l| |l| l | l| }
\hline
&  &  &  & \\
 \textbf{Pulse} & \textbf{Initial } & \textbf{Coherence} & \textbf{Final } & \textbf{Remark}\\
                      & \textbf{ State} &  & \textbf{ State} & \\
\hline
0  & 0 & 0 & 0 & - \\
0  & 1 &  0 & 1 & -\\
1  & 0 &  1 & 1 & Flips  \\
1  & 1 & 1 &  0 & Flips  \\
\hline
$\Omega(t)$   &$\vec G(t_0)$&    & \textbf{$\vec G(t)$} &  \\
\textbf{Controlled~|}   & \textbf{Target}&  \textbf{Controlled}  & \textbf{Target} &  \\
\hline
\end{tabular}
\caption{Truth table for CNOT, where $\vec G(t_0)$ and $\vec G(t)$ stand for the initial state and the final state, respectively. }
\label{table:CNOT}
\end{table}

 The proposed CNOT- gate consists of two inputs, controlled and target bit, the gate flips the state of a target bit conditional on the control bit being in the state 1. The dynamics of the system remains in its initial state if no pulse is applied. We can deduce, for instance from equation Eq.\eqref{st2}, that the final state of the system (when pulse is zero and small value of detuning) is: $(i)$ $\vec{G}\left(t\right)=(0,0,1)^T$ if it is initially prepared to be in state $|0\rangle$ or $(ii)$ $\vec{G}\left(t\right)=(0,0,-1)^T$ if it is initially prepared to be in state $|1\rangle$ (consult Eq.\eqref{gene}). To conclude if the laser pulse is \textbf{OFF}  system remains in state $|m\rangle$ assuming the system is  initially  prepared to be in state $|m\rangle$, where $m=0,1$, this mimics the first and second rows of CNOT gate as seen in Table~(\ref{table:CNOT}). But if we turn the laser pulse \textbf{ON} and if the system is initially prepared to be in state $|0\rangle$ the perturbation manages to put most of the population, i.e $\geq 0.6$, onto state $|1\rangle$, leaving few of the population in state $|0\rangle$. In such cases, the final state will have logic value \textbf{1} (because most of  the population is in state $|1\rangle$), see row three of Table~(\ref{table:CNOT}). If, however, the system is initially prepared to be in state $|1\rangle$  applying the pulse transfers most of the population to state  $|0\rangle$ thereby the corresponding logic assignment for the final state is logic value is \textbf{0} (because most of the population is in state $|0\rangle$ ), see row four of Table~(\ref{table:CNOT}). With such logic assignment we construct an operation of CNOT logic gate out of the dynamics of two level system. 
 
We state that the proposed operation is sensitive to the parameters used: detuning and effect of environment. The inclusion of noise for instance will negatively affect the coherence, consequentially the proposed implementation of CNOT gate is not viable. The use of large value of detuning far from resonance limits the analytical solution as the the Hamiltonian do not commute with itself at different times.

\subsection{Parity Checker}
\label{sec:parity}
In general, the parity of a bit string refers to the evenness or oddness of the total number of 1-bits contained in the bit string. Parity checking is commonly adopted in many digital circuits. A string of bits has 'even parity' if the number of 1-bits in the string is even, else it has odd parity. 

\begin{figure}[!htp]
\centering
\includegraphics[width=3.0 in]{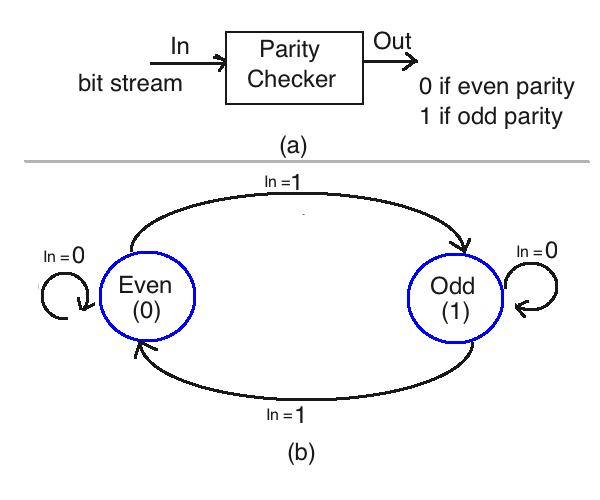}
\caption{(Color online) (a) Schematic diagram of parity checker.  (b) state diagram  of two state parity checker }
\label{parity}
\end{figure}

As shown in schematic representation of Fig.(\ref{parity}a), the circuit which has two states, even or odd, accepts a stream of bits in serial and outputs \textbf{0} if the parity so far is even and outputs \textbf{1} if odd. To ensure the total number of 1-bits is even or odd, a parity checker bit is added to bits of string. We represent these two state as circles in Fig.(\ref{parity}b), at any given moment the machine is in either one of the states, which we encode \textbf{0} for even, and \textbf{1} for odd state. We have one input bit (In=0 or 1 in the figure) that tells the machine which state to transition to. As we have one input bit, then there are two possible transition directions (shown as \textbf{0}, \textbf{1} in the figure), these inputs  show the machine which direction to go to. The output corresponding to the state is produced once the machine makes the transition, the output could be staying in the same state or move into different state. One can use either a state table or state diagram to describe the relationships between the input symbol, present state, output symbol, and next state. 
\graphicspath{{Figures//}}

For each combination of input symbol and present state, the corresponding entry provides information about the output that will be produced and the next state to which the machine will go. Each of the circle in the state diagram is a state of the machine (cf. Fig.(\ref{parity}b)). From each circle emanate directed arcs, indicating the state transitions caused by the input. These directed arc is labeled by the input symbol that causes the transition. Since both the state table and state diagram contain the same information, the choice between the two representations is a matter of convenience~\cite{kohavi2010switching}, and we use the state diagram, Fig.(\ref{parity}b), description to show the relationship between input, initial states and corresponding transitions of the parity checker.

Note that if the state of the machine is even (corresponding to logic value \textbf{0}) then input string \textbf{0} dictates to stay in the same state outputting \textbf{0} (i.e even); in like manner input string \textbf{1} means change state to odd (corresponding to logic value \textbf{1})  outputting \textbf{1} (i.e odd). On the contrary, if the state of the machine is odd (corresponding to logic value \textbf{1}) then input string \textbf{0} means stay in the same state thereby outputting \textbf{1} (i.e odd), and input string \textbf{1} means change state to even (corresponding to logic value \textbf{0}) and thus outputting \textbf{0} (i.e even). 

Corresponding to this logic machine, note that the dynamics, logic assignment as well as the variables used are the same as in preceding subsection. The present state of the logic machine, that is the state of even or odd parity, will be encoded to be the state of the pulse being \textbf{OFF} or \textbf{ON} respectively. The state of the quantum system, i.e the physical system, is encoded to represent the input of the parity checker which is \textbf{1} if the population is initially prepared to be in state $|1\rangle$ else it is \textbf{0}. In this context the term initial state refers to the state of the machine prior to the application of the input, whereas the state of the machine after the application of the input is called the final state. The next state of the machine is the final state of the the two level system-- i.e. after  interacting with the laser --state $|1\rangle$ else it is \textbf{0}. At last the output of the parity checker is encoded to be the existence and lack thereof coherence. 
\begin{table}[!h]
\centering
\begin{tabular}{ | l l| |l| l | l}
\hline
&  &  &   \\
 \textbf{Pulse} & \textbf{Initial } & \textbf{Final} & \textbf{Coherence } \\
                      & \textbf{ State} & \textbf{ State} &   \\
\hline
0  & 0 & 0 & 0  \\
0  & 1 &  1 & 0 \\
1  & 0 &  1 & 1   \\
1  & 1 & 0 &  1   \\
\hline
\textbf{PS~|}   & \textbf{Input}&  \textbf{NS}  & \textbf{Output}   \\
\hline
\end{tabular}
\caption{State transition table for parity checker, where PS is for present and NS is for next state }
\label{table:Parity}
\end{table}

If the pulse is \textbf{OFF} and the population is prepared to be in  initial state $|i\rangle, i=0,1$ then there is no change in the dynamics and therefore the system remains in its original state and no coherence is created, as we can see from Table (\ref{table:Parity}) this behavior mimics the first two rows of the parity checker.  If, however, pulse is turned \textbf{ON}, the dynamics is seen to create coherences  despite the system is initially prepared in either of the states $|i\rangle, i=0,1$ fulfilling the last column of the table.  Moreover the third column, final state of the machine is achieved by state of the physical system which is to flip state. This is to mean that if the population is initially prepared to be in the ground state, after the interaction with the pulse, most of the populations are transferred to the excited state, in like manner if the system is prepared to be on the excited state the interaction with the laser takes most of the population to the ground state, this dynamics coupled with the threshold logic assignment we introduced at the beginning of this section one is able to meet the behavior of the third column.

\section{Conclusion}
\label{sec:conclusion}
The Alhassid-Levine formalism (AL) is developed in connection to surprisal analysis while the Hioe-Eberly (HE) aimed to extend the optical Bloch equations to three and more level systems. Connection is made between these two formalisms that made use of Lie algebra to describe equation of motion for quantum system. In both approaches the Hamiltonian and the density matrix are expressed as a linear combination of the Lie group. While expansion of density matrix in terms of $SU(N)$ is true for all density matrices, the expansion of Hamiltonian in terms of the generators works only for special case of Hamiltonian. We provided a link that bridges the gap between these two formalism. With appropriate temporal profile for the laser pulse and detuning we showed how a new generator can be obtained which provides two constants of motion evolving independently on their own subspaces. Solution of the optical Bloch equation is also obtained using an approximation, i.e third order Magnus expansion, and exact analytical solution (using Wei-Norman formalism) along with the numerical solution. 

At the end before conclusion we provided examples that can be implemented by the dynamics of the two level system presented in this manuscript. We showed how the dynamics mimics the operation of CNOT gate and implementation of parity checker.

\bibliography{mybibfile}

\end{document}